\documentclass[12pt]{article}
\usepackage{mathrsfs,euscript,amsmath,epsfig,graphics,amsmath,amssymb,epsfig}

\def\co{\cos{\theta}}
\def\si{\sin{\theta}}

\def\d{\partial}
\def\l{\left(}
\def\r{\right)}
\def\la{\langle}
\def\ra{\rangle}
\def\L{{\cal L}}
\def\M{{\cal M}}
\newcommand{\be}{\begin{equation}}
\newcommand{\ee}{\end{equation}}
\newcommand{\sm}[1]{{\scriptscriptstyle \rm #1}}

\newcommand{\bg}{\begin{gather}}
\newcommand{\eg}{\end{gather}}
\newcommand{\epep}{\epsilon^{\mu\nu\lambda\rho}}
\begin{document}
\title{More about sgoldstino interpretation of HyperCP events}
\author{ S.~V.~Demidov\thanks{{\bf e-mail}: demidov@ms2.inr.ac.ru}, 
D.~S.~Gorbunov\thanks{{\bf e-mail}: gorby@ms2.inr.ac.ru}
\\
{\small{\em
Institute for Nuclear Research of the Russian Academy of Sciences, }}\\
{\small{\em
60th October Anniversary prospect 7a, Moscow 117312, Russia
}}}

\maketitle

\begin{abstract}
We further discuss possible sgoldstino interpretation of the
observation, reported by the HyperCP collaboration, of three
$\Sigma^+\to p\mu^+\mu^-$ decay events with dimuon invariant mass
$214.3$~MeV within detector resolution. With sgoldstino mass equal to
$214.3$~MeV, this interpretation can be verified at existing and
future $B$- and $\phi$-factories. We find that the most natural values
of the branching ratios of two-body $B$- and $D$-meson decays to
sgoldstino $P$ and vector meson ${\mathscr V}$ are about
$10^{-6}\div10^{-7}$. The branching ratios of $\phi$-meson decay 
$\phi\to P\gamma$ is estimated to be in the range $1.8\cdot 10^{-13}
\div 1.6\cdot 10^{-7}$, depending on the hierarchy of supersymmetry
breaking soft terms. Similar branchings for $\rho$- and
$\omega$-mesons are in the range $10^{-14}\div 3.4\cdot 10^{-7}$.  
\end{abstract}

Recent publication~\cite{HyperCP-PRL} of anomalous events 
\begin{equation}
\label{1*}
\Sigma\longrightarrow p\mu^+\mu^-\;,
\end{equation}
observed in the HyperCP experiment, renewed interest to models with
light particles whose interactions with the Standard Model fermions
violate flavor. One of the well-motivated classes of models of this
kind, supersymmetric models with light sgoldstino\footnote{See, e.g.,  
Refs. \cite{ellis,no-scale,gmm1,gmm2} for examples of models and
Refs. \cite{bhat,Brignole:1996fn,comphep} for the description of
sgoldstino interactions.}, has been suggested by the HyperCP
collaboration~\cite{HyperCP-PRL,PRESS-RELEASE} to explain the
anomalous events. The latter are three events \eqref{1*} where all
dimuon masses fall into a single bin within the detector resolution;
notably, no more decays \eqref{1*} have been observed. This
coincidence has been suggested as the first evidence for sgoldstino
production in two-body decay with subsequent decay of sgoldstino $X$
into $\mu^+\mu^-$-pair, 
\begin{equation}
\label{anomalous-decay}
\Sigma\longrightarrow pX\;,~~~X\longrightarrow \mu^+\mu^-\;.
\end{equation}
Sgoldstino mass has been fixed from measured dimuon masses as 
\begin{equation}
\label{1+}
m_X=214.3\pm0.5~{\rm MeV}\;.  
\end{equation}

It was pointed out in
Refs.~\cite{He:2005we,Deshpande:2005mb,Gorbunov:2005nu,Geng:2005ra}
that, generally, $X$-particle can be either pseudoscalar or
pseudovector, but neither scalar nor vector. Thus, only {\em 
pseudoscalar } sgoldstino (denoted by $P$ in what follows) in models
with {\em parity conserving } sgoldstino interactions (see
Ref. \cite{Gorbunov:2000cz} for a description) is 
viable~\cite{Gorbunov:2005nu}: light scalar sgoldstino, as well as
pseudoscalar sgoldstino in models with parity violating sgoldstino
interactions, cannot explain the anomalous events because of the
strong upper bounds on the widths of two-body kaon decays, $K\to\pi
X(X\to\mu^+\mu^-)$.

From the presented~\cite{HyperCP-PRL} branching ratio of the decay
\eqref{1*} (which is somewhat above the Standard Model
predictions~\cite{Bergstrom:1987wr,He:2005yn}), the following product 
has been estimated~\cite{Gorbunov:2005nu}, 
\begin{equation}
\label{2+}
|h_{12}^{(D)}|\cdot{\rm Br}^{1/2}(P\to\mu^+\mu^-)=3.8\cdot10^{-10}\;. 
\end{equation}
Here $h_{12}^{(D)}$ is sgoldstino coupling constant to pseudoscalar
current $\bar d \gamma_5 s$, 
\[
{\cal L}_{Pds} = -P\cdot(h_{12}^{(D)} \cdot \bar{d}\, i\gamma^5 s
+ \mbox{h.c.})\;,
\]
and Br$(P\to\mu^+\mu^-)$ denotes sgoldstino branching ratio to muons. 

Additional information came from the geometry of the detector and the
measured energies of muons, which enables one to put an upper limit on 
sgoldstino lifetime~\cite{Gorbunov:2005nu}, 
\begin{equation}
\label{2*}
\tau_X\lesssim 2.5\cdot10^{-11}~{\rm s}\;.
\end{equation}
Note that a lower limit~\cite{Gorbunov:2005nu} on sgoldstino lifetime
coming from the current limit on supersymmetry breaking scale is about 
\begin{equation}
\label{2a*}
\tau_X\gtrsim 1.7\cdot10^{-15}~{\rm s} 
\end{equation}
for sgoldstino mass \eqref{1+}. 

The suggested explanation of HyperCP anomalous events awaits a test in
other experiments. The processes which can be used to definitely
confirm or rule out the sgoldstino explanation have been discussed in
Ref.~\cite{Gorbunov:2005nu}. These are three-body kaon decays
\footnote{$K\to \pi\pi X$, $K\to\mu^+\mu^-$ and   $\Omega^-\to\Xi^- X$
decays have been discussed in 
Refs.~\cite{He:2005we,Deshpande:2005mb,Gorbunov:2005nu,Geng:2005ra}
as signatures to test  the hypothesis \eqref{anomalous-decay} with
some pseudoscalar or axial vector particle $X$  in a
model-independent way.} and direct sgoldstino production in $e^+e^-$
collisions. The rates of these processes are governed by the same
parameters of the sgoldstino Lagrangian, which are constrained by
Eqs.~\eqref{1+}, \eqref{2+}, \eqref{2*} and \eqref{2a*}.

The parameters are not directly constrained by the HyperCP result. 
These are other flavor-violating sgoldstino coupling constants,
$h_{jl}$ ($h_{jl}^{(U)}$, $j,l=1,2,3$, and $h_{jl}^{(D)}$, except for
$h_{12}^{(D)}$), entering 
\begin{equation}
\label{2++}
\L_{P} = -P\cdot \l h_{jl} \cdot \bar f_j\, i\gamma^5 f_l
+ \mbox{h.c.}\r\;,
\end{equation}
where $f_j$ are fermions of the Standard Model, and sgoldstino
flavor-blind couplings which do not contribute to sgoldstino width
because of kinematical constraints imposed by the small sgoldstino
mass \eqref{1+}. Nevertheless, these couplings make it possible to
extend searches for sgoldstino. Though the rates of corresponding
processes cannot be predicted because of lacking knowledge of model
parameters, one can estimate the most natural ranges for these rates
by making use of the expected patterns of MSSM soft terms. With
sgoldstino mass being fixed, these predictions can be tested
experimentally, giving an opportunity to find additional evidence for
light sgoldstino (but, generally, {\em not to compromise } the
proposed sgoldstino explanation of the HyperCP result). This issue is  
discussed in this letter.

The dominant\footnote{
The kinematically allowed decay mode $P\to e^+ e^-$ is typically
suppressed if sgoldstino mass is about 214~MeV. 
}
sgoldstino decay modes are \cite{Gorbunov:2000cz}
$P\to\mu^+\mu^- \;\; {\rm and} \;\; P\to\gamma\gamma.$
A natural range of values of the ratio of their widths is
\be
\label{br1}
\frac{\Gamma(P\to\gamma\gamma)}{\Gamma(P\to\mu^{+}\mu^{-})}
\simeq 1\div 10.
\ee
However, much larger values for this ratio, such as  $10^{4}$ are not
excluded (see \cite{Gorbunov:2005nu} for discussion). So, we are
interested in rare decays with $\mu^{+}\mu^{-}$ or $\gamma\gamma$ 
with invariant mass $214$~MeV in final states.

As promising processes, we consider two-body $B$- and $D$-meson
decays (to probe flavor-violating sector of sgoldstino couplings) and
two-body light neutral vector meson decays (to probe flavor-blind sector of
sgoldstino couplings). These searches may be performed at 
B-factories where they would cover the most natural part of the
relevant parameter space, while searches at $\phi$-factories seem less
promising, as one can probe there only a part of the relevant
parameter space leaving the rest for future experiments.

We begin with sgoldstino flavor-violating interactions \eqref{2++} in
quark sector. The coupling constants $h_{ij}$ entering these
interaction terms are determined by the ratios of the elements of
left-right part of squark squared mass matrix $\tilde m_{ij}^{(LR)2}$
to vev of auxiliary field in goldstino supermultiplet $F$, the latter
being of the order of the squared scale of supersymmetry breaking in
the complete theory, $h_{ij}=\tilde m_{ij}^{(LR)2}/(\sqrt{2}F)$. In
particular, non-zero off-diagonal entries of squark squared mass
matrix lead to two-body  pseudoscalar $B$- and $D$-meson decays to a
vector meson ${\mathscr V}$ and pseudoscalar sgoldstino\footnote{Two-body decays
of pseudoscalar mesons into pseudoscalar mesons and pseudoscalar
sgoldstino are strongly suppressed by parity conservation in
sgoldstino interactions.},  
\begin{equation}
\label{4+}
P_{B,D}\to {\mathscr V} P\;.
\end{equation}
Let $p$, $p_{\sm {\mathscr V}}$ be momenta of $P_{B,D}$ and ${\mathscr V}$,
respectively, $q\equiv p-p_{\sm {\mathscr V}}$, and
$\epsilon_{\mathscr V}$ be the
polarization vector of ${\mathscr V}$. With these notations, the hadronic matrix
elements
\[
O_{P_{B,D},{\mathscr V}}\l q^2\r\equiv \la {\mathscr V}(p_{\sm
  {\mathscr V}},\epsilon_{\sm {\mathscr V}})|\bar q_j\gamma^5 q_l|P_{B,D}(p)\ra
\]
 entering the decay amplitudes
${\cal M}\l P_{B,D}\to {\mathscr V} P \r=h_{jl}\cdot P\l q\r
\cdot O_{P_{B,D},{\mathscr V}}\l q^2\r$
(where $P(q)$ is the wave function of the outgoing sgoldstino) 
can be expressed via experimentally measured and/or theoretically
predicted form-factors $A_0^{(P_{B,D},{\mathscr V})}(q^2)$ as follows (see, e.g.,
Ref.~\cite{Cheng:2003sm} for notations and other details), 
\begin{equation}
\label{5+}
O_{P_{B,D},{\mathscr V}}\l q^2\r=A_0^{(P_{B,D},{\mathscr
    V})}(q^2)\cdot \frac{-2i m_{\mathscr V}}{m_j+m_l}\;,
\end{equation}
where $m_{\mathscr V}$ and $m_i$ are masses of ${\mathscr V}$ and quark $q_i$, respectively;
we use $m_b=4.8$~GeV, $m_c=1.35$~GeV,
$m_s=0.13$~GeV~\cite{Cheng:2003sm} in our estimates. Making use of the 
expression \eqref{5+}, one finds the partial width 
\begin{equation}
\label{5*}
\Gamma\l P_{B,D}\to {\mathscr V} P \r  =
\frac{|h_{jl}|^2}{16\pi}\l A_0^{(P_{B,D},{\mathscr V})}(m_P^2)\r^2 
\cdot
\frac{m_{P_{B,D}}^3\lambda^3(m_{P_{B,D}},m_{\mathscr V},m_P)}{\l
m_j+m_l\r^2}\;,
\end{equation}
and
$ \lambda(m_1,m_2,m_3) =\sqrt{\l
1-\frac{\l m_2+m_3\r^2}{m_1^2}\r\l 1-\frac{\l
m_2-m_3\r^2}{m_1^2}\r}\;$.
The values of the form factors $A_0^{(P_{B,D},{\mathscr V})}$ entering Eq. \eqref{5*} can
be found in
literature~\cite{Melikhov:2000yu,Ebert:2003cn,Aliev:2006vs,Cheng:2003sm}. 

To illustrate the level of precision required to probe the models 
consistent with HyperCP events, we present estimates of the widths in
three different types of supersymmetric models. Two of them are
\begin{align*}
\mbox{model I}\; & : \; \; \; h_{jl}\sim h_{12}^{(D)}\;,\;\; j \ne l \\
\mbox{model II}\; & : \; \; \; h_{jl}\sim \frac{A}{F}\cdot
max(m_j,m_l)\;, 
\end{align*}
where $A$ is a flavor-independent constant. The third model is a 
concrete phenomenologically viable example of left-right
supersymmetric model, where parity conservation in sgoldstino 
interactions is guaranteed~\cite{Gorbunov:2000cz}. In the first model
all off-diagonal entries in squark squared mass matrix are  of the
same order. In the second model there is a hierarchy in matrix
elements $h_{jl}$ reflecting the hierarchy of quark masses. The latter
situation  is more realistic, at it is typical for minimal
supersymmetric models like mSUGRA or models with gauge mediation of   
supersymmetry breaking, where soft supersymmetry breaking trilinear
terms are proportional to the Yukawa matrix. For the  model III we
take a general left-right SUSY model of the type presented in
Ref.~\cite{Gorbunov:2000cz}, with left-right symmetry broken at energy 
scale $M_{R}= 3$~TeV. We assume (see notations in 
Ref.~\cite{Gorbunov:2000cz}) universality at the SUSY breaking scale  
$\sqrt{F}= 30$~TeV, {\it i.e.} ${\mathbf   A}^{(i)}= {\mathbf
 Y}^{(i)}A^{(i)}$, choose $A^{(1)} = A^{(2)} = M_{1/2}$, $\tan{\beta}
= 3.6$ and neglect mixing between Higgs doublets in
doublet-doublet splitting \cite{Mohapatra:1996vg}. The left-right
entries in the squark squared mass matrices $\tilde{m}_{D(U)}^{LR\;2}$ are
obtained by making use  of one-loop renormalization group equations 
for gauge, Yukawa, soft trilinear coupling constants and gaugino
masses~\cite{Setzer:2005hg}. The value of $A^{(1)}$ at $M_{R}=
3$~TeV is fixed by \eqref{2+}.  Under our assumptions, all relevant
parameters in model III are then completely determined.

For these three models it is straightforward to estimate the partial 
widths of the processes \eqref{4+} with  real sgoldstino decaying into
$\mu^+\mu^-$, $P_{B,D}\to {\mathscr V} P(P\to\mu^+\mu^-)\;.$ The results are
summarized in Table \ref{table-1}.
\begin{table*}[htb!]
{\renewcommand{\arraystretch}{1.0}
\centering
\begin{minipage}{\textwidth}
\centering
\begin{tabular}{|c|c|c|c|c|c|}
\hline
decay & $h_{jl}$ &
$A_0^{(P_{B,D},{\mathscr V})}$ & ${\rm Br}_{({\rm model\; I})}$ & ${\rm
Br}_{({\rm model\; II})}$ 
& ${\rm Br}_{({\rm model\; III})}$
\\
\hline
$B_s\to \phi P(P\to\mu^+\mu^-)$ & $h^{(D)}_{23}$ &
$0.42$\;\cite{Melikhov:2000yu} & $6.5\cdot 10^{-9}$ &
$8.8\cdot 10^{-6}$ & $8.7\cdot 10^{-6}$ \\
$B_s\to K^{*0} P(P\to\mu^+\mu^-)$ & $h^{(D)}_{13}$ &
$0.37$\;\cite{Melikhov:2000yu} & 
$5.3\cdot 10^{-9}$ & $7.2\cdot 10^{-6}$ & $2.3\cdot
10^{-7}$\\
\hline
$B_c^+\to D^{*+} P(P\to\mu^+\mu^-)$ & $h_{13}^{(D)}$ & $0.14$\;\cite{Ebert:2003cn} &
$3.2\cdot10^{-10}$ & $4.4\cdot 10^{-7}$ & $1.4\cdot 10^{-8}$ \\
$B_c^+\to D_s^{*+} P(P\to\mu^+\mu^-)$ & $h^{(D)}_{23}$ &
$0.14$\footnote{
We did not find any estimate of this formfactor in literature and use
this value as an order-of-magnitude estimate,  which is sufficient for
our study.} &
$3.0\cdot10^{-10}$ & $4.0\cdot 10^{-7}$ & $4.0 \cdot 10^{-7}$ \\
$B_c^+\to B^{*+} P(P\to\mu^+\mu^-)$ & $h_{12}^{(U)}$ & $0.23$\;\cite{Aliev:2006vs}&
$4.1\cdot10^{-10}$ & $4.4\cdot 10^{-8}$ & $8.2\cdot 10^{-7}$ \\
\hline
$B^+\to K^{*+} P(P\to\mu^+\mu^-)$ & $h^{(D)}_{23}$ &
$0.31$\;\cite{Cheng:2003sm}& $3.8\cdot 10^{-9}$
& $5.2\cdot 10^{-6}$ & $5.1\cdot 10^{-6}$\\
$B^0\to K^{*0} P(P\to\mu^+\mu^-)$ & & &
$3.5\cdot 10^{-9}$ & $4.8\cdot 10^{-6}$ & $4.7\cdot 10^{-6}$\\
\hline
$B^0\to \rho P(P\to\mu^+\mu^-)$ & $h_{13}^{(D)} $ &
$0.28$\;\cite{Cheng:2003sm} & $3.1\cdot 10^{-9}$ &
$4.2\cdot 10^{-6}$ & $1.4\cdot 10^{-7}$\\
$B^+\to \rho^+ P(P\to\mu^+\mu^-)$ & & & $3.3\cdot 10^{-9}$
& $4.6\cdot 10^{-6}$ & $1.3\cdot 10^{-7}$ \\
\hline
$D^0\to \rho P(P\to\mu^+\mu^-)$ & $h^{(U)}_{12}$ &
$0.64$\;\cite{Cheng:2003sm} & $1.4\cdot 10^{-9}$ &
$1.5\cdot 10^{-7}$ & $2.8\cdot 10^{-6}$ \\
$D^+\to \rho^+ P(P\to\mu^+\mu^-)$ & & & $3.5\cdot 10^{-9}$&
$3.7\cdot 10^{-7}$ & $7.0\cdot 10^{-6}$ \\
\hline
\end{tabular}
\end{minipage}}
\caption{Branching ratios of decays
$P_{B,D}\to {\mathscr V}P(P\to\mu^+\mu^-)$ in the models I, II and III. Branching
ratios of decays $P_{B,D}\to {\mathscr V} P(P\to\gamma\gamma)$ are given by
the same numbers multiplied by
$\Gamma(P\to\gamma\gamma)/\Gamma(P\to\mu^+\mu^-)$.  
\label{table-1}
}
\end{table*}
For the widths of similar processes, but with sgoldstino decaying into
photons, $P_{B,D}\to {\mathscr V} P(P\to\gamma\gamma)\;,$ one gets the same
numbers multiplied by the ratio 
$\Gamma(P\to\gamma\gamma)/\Gamma(P\to\mu^+\mu^-)$, whose estimates are
given in \eqref{br1}. 

Comparing the results presented in Table~\ref{table-1} with statistics
of $B$- and $D$-meson decays collected by B-factories, one concludes
that both $\mu^+\mu^-$ and $\gamma\gamma$ decay channels can be probed
for the most natural choice of parameters (models II, III). Moreover,
a part of expected region for $h_{23}^{(D)}$ is already excluded by
the results~\cite{Eidelman:2004wy}   
\begin{equation}
\label{exp-exc}
\begin{split}
{\rm Br}\l B^+\to K^{*+}\mu^+\mu^- \r&<2.2\cdot10^{-6}\;,\\
{\rm Br} \l B^0\to K^{*0} \mu^+\mu^-\r &=\l 1.3\pm 0.4\r\cdot10^{-6}\;.
\end{split}
\end{equation}
Study of the model I requires, generally, higher statistics. At
B-factories model I could be probed if sgoldstino decay mode into
photons dominates by one to two orders of magnitude over $\mu^+\mu^-$
mode. 

It is worth noting that coupling constants $h_{13}^{(D)}$,
$h_{23}^{(D)}$ and $h_{12}^{(U)}$ determine the rates of several
decays each.  Hence, the ratios of the corresponding rates do not
depend on the values of these couplings. This would allow for
independent check of sgoldstino interpretation, would any of the
anomalous decays listed in Table~\ref{table-1} be observed. In
particular, Eq.~\eqref{exp-exc} implies that branching ratios of
$B_s\to \phi P(P\to\mu^+\mu^-)$ and $B_c^+\to D_s^{*+}
P(P\to\mu^+\mu^-)$ have to be at least three times smaller than the
numbers in the two last columns of Table~\ref{table-1}. 

Another signature of models with light sgoldstino we would like to
discuss is neutral vector meson ${\mathscr V}$ decays to sgoldstino
and photon, 
\begin{equation}
\label{8+}
{\mathscr V}\to P\gamma.
\end{equation}
The most sensitive are the decays of $\phi$-, $\omega$- and
$\rho$-mesons, whose rates are almost saturated by sgoldstino
couplings to gluons, as we find below.  For $J/\psi$- and $\Upsilon$-mesons
the corresponding branching ratios do not exceed $10^{-9}$ (see Table
5 in Ref.~\cite{Gorbunov:2000th}). 
\begin{figure}[htb!]
\begin{center}
\includegraphics[width=0.8\columnwidth]{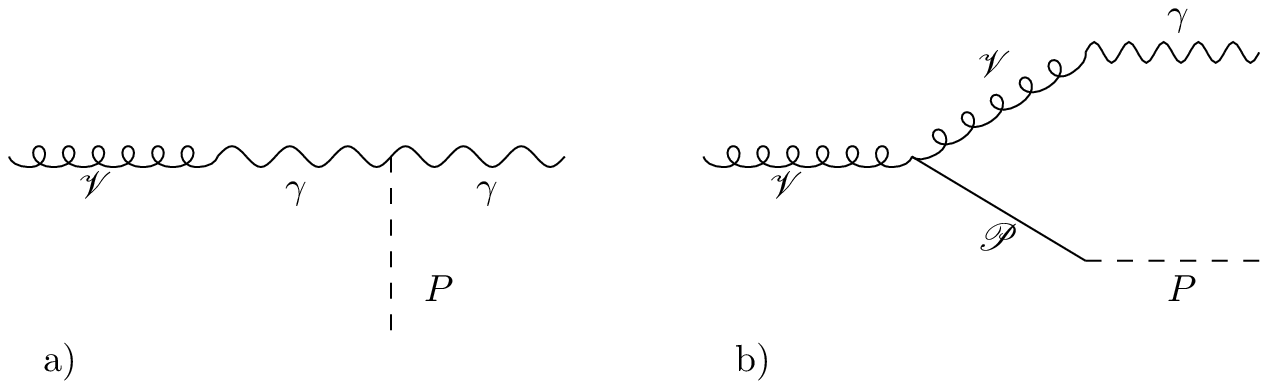}
\end{center}
\caption{\label{diagr}
Diagrams contributing to the decay ${\mathscr V}\to\gamma P$.
}
\end{figure}

There are two different contributions, $\M_1$ and $\M_2$, to the
amplitude of the decay \eqref{8+}. These are due to sgoldstino
couplings to photons and to gluons, 
\begin{equation}
\label{9+} 
\L_1=
\frac{1}{4\sqrt{2}}\epsilon^{\mu\nu\lambda\rho}\left(\frac{M_{\gamma\gamma}}{F}
PF_{\mu\nu}F_{\lambda\rho} +
\frac{M_3}{F} PG^a_{\mu\nu}G^a_{\lambda\rho}\right)
\;,~~
\end{equation}
(where $M_{\gamma\gamma}=M_1 \cos^2\theta_{\sm W}+M_2\sin^2\theta_{\sm
W} $ and $M_1$, $M_2$ and $M_3$ are masses of $U(1)_{\sm Y}$-,
$SU(2)_{\sm W}$- and $SU(3)_{c}-$gauginos, respectively)
and sgoldstino flavor-conserving interactions with quarks
(\ref{2++}). 

The relevant interactions of vector meson ${\mathscr V}$ are 
${\mathscr V}-\gamma$ mixing~\cite{Meissner:1987ge}, 
\begin{equation}
\label{9++} 
\L_{{\mathscr V},1}=
\frac{e}{g}a_{\mathscr V}m_{\mathscr V}^{2}A_\mu{\mathscr V}_{\mu}\;,~~~g\simeq 8.6 
\end{equation}  
(where $m_{\mathscr V}$ is the mass of the corresponding vector meson
and $\left(a_{\rho},\;a_{\omega},\;a_{\phi}\right) =
\left(\sqrt{2},\;\frac{\sqrt{2}}{3},\;-\frac{2}{3}\right)$), and
coupling to photon and pseudoscalar mesons ${\mathscr
P}$~\cite{Escribano:2005qq},    
\begin{eqnarray}
\label{9**} 
{\mathcal L}_{{\mathscr V},2} = 
eg_{{\mathscr V}{\mathscr P}
\gamma}\epsilon_{\mu\nu\lambda\rho}\d_{\mu}A_{\nu}\d_{\lambda}{\mathscr V}_{\rho}{\mathscr P}\;,
\end{eqnarray}
where the values of coupling constants $g_{{\mathscr V}{\mathscr
P}\gamma}$ are tuned to saturate the ${\mathscr P}\to {\mathscr V}\gamma$ 
or ${\mathscr V}\to{\mathscr P}\gamma$ decay rates and are presented
in Ref.~\cite{Escribano:2005qq}. 

Couplings \eqref{9+} and \eqref{9++} give rise to the following
contribution to the decay amplitude 
\[
\M_1=\epsilon_{\mu}(p)\cdot i\frac{e}{g}a_{\mathscr V}
\cdot\frac{-i}{m_{\mathscr V}^{2}}\cdot \l
-i\sqrt{2}\frac{M_{\gamma\gamma}}{F} \r \cdot
\epsilon^{\mu\nu\lambda\rho} p_{\lambda}k_{\rho}
\epsilon^{*}_{\nu}(k)\;, 
\]
where $p$ and $k$ stand for momenta of ${\mathscr V}$-meson and $\gamma$, 
respectively, while $\epsilon(p)$ and $\epsilon(k)$ denote their
polarizations. This contribution is illustrated in Fig.\ref{diagr}a.  

The interactions \eqref{9+} give rise to $P-{\mathscr P}$
mixing~\cite{Gorbunov:2000th}, 
\begin{equation}
\label{10+}
\L_{P{\mathscr P}}=C_{P{\mathscr P}}\cdot P{\mathscr P}\;,\\
\end{equation}
where
\[
C_{P\pi^{0}}= -\frac{\sqrt{2}\pi m_{\pi}^{2}M_3}{3\alpha_{s}(M_{3})F}\frac{m_{u} -
m_{d}}{m_{u} + m_{d}}f_{\pi}\;,
\]
\[
C_{P\eta (\eta^{\prime})} 
= \frac{\pi M_3 m_{\eta
(\eta^{\prime})}^{2}}{\sqrt{3}\alpha_{s}F}f_{\eta (\eta^{\prime})} 
+ \frac{Bf_{\pi}Am_{s}}{\sqrt{6}F}\left(\sqrt{2}\co  \pm \si\right),
\]
\[
f_{\eta (\eta^{\prime})} = f_{8}\cos{\theta_{8}} \mp
\sqrt{2}f_{0}\sin{\theta_{0}},
\]
where upper sign stands for $\eta$-meson and lower for $\eta^{\prime}$. Here
$B_0\approx m_K^2/m_s\simeq 2$~GeV and we neglect small ($\sim
10^{-2}$) mixing between $\pi^{0}$ and $\eta,\;\eta^{\prime}$-mesons,
$\eta - \eta^{\prime}$ mixing angle $\theta = -15.4^{\circ}$ and we
adopt two mixing angle scheme for the parametrization of the gluonic
matrix elements~\cite{Escribano:2005qq},\cite{Gross:1979ur}  
\begin{equation}
\label{1.1}
\la
\eta
|
\frac{3\alpha_{s}(M_{3})}{4\pi}G_{\mu\nu}^{a}\tilde{G}_{\mu\nu}^{a}
|0\ra = \sqrt{\frac{3}{2}}m_{\eta (\eta^{\prime})}^{2}f_{\eta (\eta^{\prime})}
\end{equation}
\[
\la \pi^{0} |
\frac{3\alpha_{s}(M_{3})}{4\pi}G_{\mu\nu}^{a}\tilde{G}_{\mu\nu}^{a}
|0\ra = -\frac{m_{u} - m_{d}}{m_{u} + m_{d}}f_{\pi}m_{\pi}^{2}\;. 
\]
where we use $f_{0} = 1.17f_{\pi}$, $f_{8} = 1.26f_{\pi}$ and
$\theta_{8} = -21.2^{\circ}$, $\theta_{0} = -9.2^{\circ}$
\cite{Feldmann:2002kz}. 

Couplings \eqref{9**} and \eqref{10+} give rise to the contribution
\[
\M_2= \sum_{\mathscr P}\left(-ig_{{\mathscr V}{\mathscr P}\gamma}\right)
\epsilon_{\mu}\left(p\right) \epep p_{\lambda}k_{\rho}
\epsilon^{*}_{\nu}(k)
\frac{i}{m_{P}^{2} - m_{\eta}^{2}}\cdot iC_{P\eta}
\]
illustrated by the diagram shown in Fig.\ref{diagr}b. Here sum runs
over $\pi^{0},\;\eta$ and $\eta^{\prime}$ mesons.

Finally, the rate of the decay ${\mathscr V}\to
P\gamma$ in the case  of unpolarized vector meson reads 
\begin{align*}
&\Gamma\l {\mathscr V}\to P\gamma\r=
\frac{C_{\mathscr V}^2}{96\pi}\frac{\l m_{\mathscr
V}^2-m_P^2\r^3}{m_{\mathscr V}^3}\;,
\end{align*}
\begin{equation}
C_{\mathscr V} = -\frac{\sqrt{2}a_{\mathscr V}eM_{\gamma\gamma}}{gF}
+ \sum_{{\mathscr P} = \pi^{0},\eta,\eta^{\prime}}
\frac{eg_{{\mathscr V}{\mathscr P}\gamma}C_{P{\mathscr P}}}
{m_{P}^{2} - m_{\mathscr P}^{2}}
\end{equation}
In most models, the gluino mass is large, so that the dominant
contribution comes from sgoldstino couplings to gluons. The
corresponding coupling constant is proportional to the ratio $M_3/F$.
It reaches the maximal value in the unitarity limit $M_3\simeq
\sqrt{F}$ at the smallest value of $\sqrt{F}$, which is consistent
with current bounds on the scale of supersymmetry breaking,
$\sqrt{F}\gtrsim 500$~GeV. So, we have 
\begin{equation}
\label{11-add++}
\left[\frac{M_3}{F}\right]_{max}\sim 2\cdot 10^{-3}~{\rm GeV}^{-1}\;.
\end{equation}
If $M_3/F$ is much smaller than this value, the dominant contribution
to $C_\phi$ comes from either $M_{\gamma\gamma}/F$ or $A/F$ terms, at
least one of which is bounded from below by the upper limit on
sgoldstino lifetime \eqref{2*},  
\begin{equation}
\label{11-add**} 
\left[\frac{M_{\gamma\gamma}}{F},\frac{\left| A\right|}{F}\right]_{min}
\sim 1.5\cdot 10^{-5}~{\rm GeV}^{-1}\;.
\end{equation}
When considering the case of small $M_{3}$, we present the numerical
results for the case $\left| A\right|\ll M_{\gamma\gamma}$. In the
opposite case one obtains about two orders of magnitude smaller
numbers for the lower bounds of the intervals. 

Thus, the branching ratios of ${\mathscr V}\to P\gamma$ decays are
expected to be within the intervals
\begin{eqnarray}
\label{11+}
1.8\cdot 10^{-13} < Br_\phi < 1.6\cdot 10^{-7} \\
9.1\cdot 10^{-15} < Br_\rho < 3.3\cdot 10^{-7}\\
1.8\cdot 10^{-14} < Br_\omega < 3.4\cdot 10^{-7}
\end{eqnarray} 
The produced sgoldstino subsequently decays either into photons or
into muons. Comparing the values of the branching ratios in the
 interval \eqref{11+} to the collected world statistics of
$\phi$-mesons (about a few billion) one concludes that the part of
the interval \eqref{11+} with the largest branchings may be tested
with existing experimental data at $\phi$-factories, while the large
part could be tested only with significant increase in statistics. 
Note that observation of two-body decays of different vector mesons
would offer an opportunity to measure all relevant parameters
$M_{\gamma\gamma}$, $A$, $M_3$, $F$.  

To conclude, by studying two-body decays of $B$-, $D$-, $\phi$-,
$\omega$- and $\rho$-mesons one can probe the models with
light sgoldstino, capable of explaining HyperCP anomalous events. 
Currently available statistics is sufficient to probe the most natural
left-right symmetric models and models with the hierarchy in 
sgoldstino flavor-violating couplings similar to the hierarchy of
quark masses. The rest of models awaits larger statistics. The same
conclusion holds for quite similar baryon two-body decay modes like
$\Omega^-\to\Xi^-P$~\cite{He:2005we,Deshpande:2005mb,Gorbunov:2005nu},
$\Omega^0_c\to\Xi^0_cP$, etc.: though the rates of (some of) these
decays can be predicted from the HyperCP data, the current statistics
is too low, so that future experiments are needed to test the
sgoldstino explanation by searching for these decays.

If observed, these decays would not only confirm the sgoldstino
interpretation of HyperCP results (which have been claimed as the first
evidence for supersymmetry~\cite{PRESS-RELEASE}) but would give an 
opportunity to measure sgoldstino life-time and to probe the pattern
of sgoldstino flavor violating couplings to quarks. The former would
fix the ratios of MSSM soft terms to the squared scale of 
supersymmetry breaking, while the latter would enable one to estimate
the values of off-diagonal entries in squarks squared mass matrix,
which can be tested, in turn, by searching for FCNC processes. 

The negative results of searches for sgoldstino in two-body $B$- and
$D$-meson decays at the level of branchings as high as $10^{-7}\div
10^{-8}$ (models II and III) would imply either non-sgoldstino explanation of
the HyperCP results or fairly special pattern of off-diagonal squark
masses, atypical for simple supersymmetric extensions of the Standard
Model.  Hence, some mechanism (e.g., additional flavor symmetry)
should have to work in that case to provide for this pattern. 

We are indebted to A.~Bondar, M.~Danilov, S.~Eidelman, P.~Pakhlov,
V.~Rubakov for numerous helpful discussions. This work was supported
in part by grant NS-7293.2006.2 (government contract
02.445.11.7370), by RFBR grant 05-02-1736 and by the Dynasty
Foundation (awarded by the Scientific board of ICFPM); the work of
D.G.\ was supported in part by the grant RFBR-04-02-17448. D.G.\
thanks CERN Theory Division, where part of this work was done, 
for hospitality.


\begin{thebibliography}{20}

\bibitem{HyperCP-PRL} H.~Park {\it et al.}  [HyperCP Collaboration],
Phys.\ Rev.\ Lett.\  {\bf 94}, 021801 (2005).

\bibitem{ellis}
J.~R.~Ellis, K.~Enqvist and D.~V.~Nanopoulos,
Phys.\ Lett.\ B {\bf 147} 99 (1984),
J.~R.~Ellis, K.~Enqvist and D.~V.~Nanopoulos,
Phys.\ Lett.\ B {\bf 151} 357 (1985).

\bibitem{no-scale}
T.~Bhattacharya and P.~Roy,
Phys.\ Rev.\ D {\bf 38} 2284 (1988).

\bibitem{gmm1} 
G.~F.~Giudice and R.~Rattazzi,
Phys.\ Rept.\  {\bf 322} 419 (1999)
[arXiv:hep-ph/9801271].

\bibitem{gmm2} 
S.~L.~Dubovsky, D.~S.~Gorbunov and S.~V.~Troitsky,
Phys.\ Usp.\  {\bf 42}, 623 (1999)
[Usp.\ Fiz.\ Nauk {\bf 169}, 705 (1999)]. 

\bibitem{bhat}
T.~Bhattacharya and P.~Roy,
Phys.\ Lett.\ B {\bf 206} 655 (1988).

\bibitem{Brignole:1996fn}
A.~Brignole, F.~Feruglio and F.~Zwirner,
Nucl.\ Phys.\ B {\bf 501} 332 (1997). 

\bibitem{comphep} D.~S.~Gorbunov and A.~V.~Semenov, 
``CompHEP package with light gravitino and sgoldstinos,'' 
arXiv:hep-ph/0111291.

\bibitem{PRESS-RELEASE} {\tt http://www.fnal.gov/pub/news05/HyperCP.html}

\bibitem{He:2005we}
X.~G.~He, J.~Tandean and G.~Valencia,
Phys.\ Lett.\ B {\bf 631} 100 (2005)
[arXiv:hep-ph/0509041].

\bibitem{Deshpande:2005mb}
N.~G.~Deshpande, G.~Eilam and J.~Jiang,
Phys.\ Lett.\ B {\bf 632}  212 (2006)
[arXiv:hep-ph/0509081].

\bibitem{Gorbunov:2005nu}
D.~S.~Gorbunov and V.~A.~Rubakov,
Phys.\ Rev.\ D {\bf 73}, 035002 (2006)

\bibitem{Geng:2005ra}
C.~Q.~Geng and Y.~K.~Hsiao,
Phys.\ Lett.\ B {\bf 632} 215 (2006)

\bibitem{Gorbunov:2000cz}
D.~S.~Gorbunov and V.~A.~Rubakov, 
Phys.\ Rev.\ D {\bf 64}, 054008 (2001).

\bibitem{Bergstrom:1987wr}
L.~Bergstrom, R.~Safadi and P.~Singer,
Z.\ Phys.\ C {\bf 37} 281 (1988).

\bibitem{He:2005yn}
X.~G.~He, J.~Tandean and G.~Valencia,
Phys.\ Rev.\ D {\bf 72} 074003 (2005)

\bibitem{Cheng:2003sm}
H.~Y.~Cheng, C.~K.~Chua and C.~W.~Hwang,
Phys.\ Rev.\ D {\bf 69} 074025 (2004)
[arXiv:hep-ph/0310359].

\bibitem{Melikhov:2000yu}
D.~Melikhov and B.~Stech,
Phys.\ Rev.\ D {\bf 62} 014006 (2000)
[arXiv:hep-ph/0001113].

\bibitem{Ebert:2003cn}
D.~Ebert, R.~N.~Faustov and V.~O.~Galkin,
Phys.\ Rev.\ D {\bf 68} 094020 (2003)
[arXiv:hep-ph/0306306].

\bibitem{Mohapatra:1996vg}
R.~N.~Mohapatra and A.~Rasin,
Phys.\ Rev.\ D {\bf 54} 5835 (1996) 5835
[arXiv:hep-ph/9604445].

\bibitem{Aliev:2006vs}
T.~M.~Aliev and M.~Savci,
arXiv:hep-ph/0601267.

\bibitem{Setzer:2005hg}
N.~Setzer and S.~Spinner,
Phys.\ Rev.\ D {\bf 71} 115010 (2005)

\bibitem{Eidelman:2004wy}
S.~Eidelman {\it et al.}  [Particle Data Group],
Phys.\ Lett.\ B {\bf 592}, 1 (2004).

\bibitem{Meissner:1987ge}
See, e.g., U.~G.~Meissner,
Phys.\ Rept.\  {\bf 161} 213 (1988).

\bibitem{Escribano:2005qq}
P.~Ball, J.~M.~Frere and M.~Tytgat,
Phys.\ Lett.\ B {\bf 365}  367 (1996)
R.~Escribano and J.~M.~Frere,
JHEP {\bf 0506}  029 (2005)

\bibitem{Feldmann:2002kz}
T.~Feldmann and P.~Kroll,
Phys.\ Scripta {\bf T99} (2002) 13

\bibitem{Gorbunov:2000th}
D.~S.~Gorbunov, 
Nucl.\ Phys.\ B {\bf 602}, 213 (2001).

\bibitem{Gross:1979ur}
D.~J.~Gross, S.~B.~Treiman and F.~Wilczek,
Phys.\ Rev.\ D {\bf 19} 2188 (1979).
\end{thebibliography}
\end{document}